\documentclass[12pt]{article}
\usepackage{cite,amssymb,amsmath,graphicx,graphics}
\usepackage{color}
\usepackage[colorlinks=true,linkcolor=red]{hyperref}
\usepackage{hyperref}
\begin{document}
\title{\bf The 5D Standing Wave Braneworld With Real Scalar Field}
\author{{\bf Merab Gogberashvili}\\
Andronikashvili Institute of Physics,\\
6 Tamarashvili Street, Tbilisi 0177, Georgia \\
and \\
Javakhishvili State University,\\
3 Chavchavadze Avenue, Tbilisi 0128, Georgia\\
{\sl E-mail: gogber@gmail.com}\\\\
{\bf Pavle Midodashvili}\\
Ilia State University,\\
3/5 Cholokashvili Avenue, Tbilisi 0162, Georgia\\
{\sl E-mail: pmidodashvili@yahoo.com}
}
\maketitle
\begin{abstract}
We introduce the new 5D braneworld with the real scalar field in the bulk. The model represents the brane which bounds collective oscillations of gravitational and scalar field standing waves. These waves are out of phase, i.e. the energy of oscillations passes back and forth between the scalar and gravitational waves. When the amplitude of the standing waves is small the brane width and the size of the horizon in extra space are of a same order of magnitude, and matter fields are localized in extra dimension due to the presence of the horizon. When oscillations are large trapping of matter fields on the brane is provided mainly by the pressure of bulk waves. It is shown that in this case the mass of the lightest KK mode is determined by the smaller energy scale corresponding to the horizon size, i.e. these modes can be created in accelerators at relatively low energies, which gives a chance to check the present model.
\end{abstract}

\vskip 0.5cm

\section{Introduction}

Braneworld models involving large extra dimensions \cite{Hi-1,Hi-2,brane-1,brane-2,brane-3,brane-4} have been very useful in addressing several open questions in high energy physics (e.g. the hierarchy problem and the nature of flavor). Most of the braneworlds are realized as time-independent field configurations. However, mostly within the framework of cosmological studies, there were proposed models with time-dependent metrics and matter fields, as well as branes with tensions varying in time \cite{S-1,S-2,S-3,S-4}. One of the 5D braneworld models with non-stationary metric coefficients was proposed by one of us in \cite{Wave-1,Wave-2,Wave-3,Wave-4} (for the generalization to 6D case see \cite{6D-1,6D-2}). In this scenario the braneworld was generated by standing gravitational waves coupled to a phantom-like bulk scalar field, rapid oscillations of these waves provide universal gravitational trapping of zero modes of all kinds of matter fields on the brane \cite{Loc-1,Loc-2,Loc-3,Loc-4,Loc-5}.

In this paper we introduce the new non-stationary 5D braneworld generated by standing waves of the gravitational and real scalar fields, instead of the phantom-like scalar fields of \cite{Wave-1,Wave-2,Wave-3,Wave-4}. The model also has new features: it does not require a bulk cosmological constant and the oscillation frequency of the standing waves can be small.


\section{The Model}

We consider 5D space-time without bulk cosmological constant containing a brane and a non-self interacting real scalar field coupled to gravity:
\begin{equation}\label{action}
S = \int d^5x \sqrt g \left( \frac{M^3}{2}R + \frac 12 g^{MN}\partial _M \varphi \partial _N\varphi + L_B \right)~.
\end{equation}
Here $L_B$ is the brane Lagrangian and $M$ is the 5D fundamental scale, which relates to the 5D Newton constant, $G=1/(8\pi M^3)$. Capital Latin indexes numerate the coordinates of 5D space-time with the signature $(+,-,-,-,-)$, and we use the units where $c = \hbar = 1$. Variation of the action (\ref{action}) with respect to $g_{AB}$ leads to the 5D Einstein equations:
\begin{equation}\label{EinsteinEquations1}
R_{AB} - \frac 12 g_{AB}R = \frac {1}{M^3}\left(\sigma_{AB} + T_{AB}\right)~.
\end{equation}
Here the source term is the combination of the energy-momentum tensors of the brane, $\sigma_{AB}$, and of the bulk scalar field,
\begin{equation}\label{ScalarFieldEnergyMomentumTensor}
T_{AB} = \partial _A\varphi \partial _B\varphi - \frac 12 g_{AB} \partial ^C\varphi \partial _C\varphi~.
\end{equation}
Using (\ref{ScalarFieldEnergyMomentumTensor}) the Einstein equations (\ref{EinsteinEquations1}) can be rewritten in the form:
\begin{equation}\label{EinsteinEquations2}
R_{AB}= \frac {1}{M^3}\left(\sigma_{AB}-\frac{1}{3}g_{AB}\sigma + \partial_A \varphi \partial_B \varphi \right)~,
\end{equation}
where $\sigma = g^{AB}\sigma_{AB}$.

To solve the equations (\ref{EinsteinEquations2}) we take the metric {\it ansatz}:
\begin{equation} \label{MetricAnsatz}
ds^2 = \frac {e^S}{(1 - a|r|)^{2/3}}\left( dt^2 - dr^2 \right) - (1 - a|r|)^{2/3}\left( e^u dx^2 + e^u dy^2 + e^{-2u}dz^2 \right)~,
\end{equation}
where $a$ is a positive constant and $S=S(|r|)$, $u=u(t,|r|)$ are some functions.

The metric (\ref{MetricAnsatz}) is some combination of the 5D generalizations of the known domain wall solution \cite{domain-1,domain-2,domain-3} (when $S=u=0$) and of the colliding plane wave solutions \cite{plane-1,plane-2,plane-3} (when $S=a=0$). Similar {\it ansatz} in the 4D case was considered by one of us in \cite{GMS}.

To find a solution to the system of Einstein and background real scalar field equations let us assume that the 5D scalar field,
\begin{equation}
\varphi \equiv \varphi (t, |r|)~,
\end{equation}
depends only on time, $t$, and on absolute value of the extra coordinate, $|r|$. Then its equation of motion:
\begin{equation}\label{phi}
\frac{1}{\sqrt g}\partial _M\left( \sqrt g g^{MN}\partial _N\varphi \right) = 0~,
\end{equation}
where
\begin{equation}\label{DeterminantOfMetric}
\sqrt g  = (1 - a|r|)^{1/3}e^S~,
\end{equation}
is the determinant of the background metric (\ref{MetricAnsatz}), reduces to
\begin{equation}\label{BulkScalarFieldEquation}
\varphi'' + \left[2\delta (r) - \frac{a}{1 - a|r|} \right]\varphi' - \ddot \varphi = 0~,
\end{equation}
where overdots and primes denote derivatives with respect to $t$ and $|r|$, respectively.

For the {\it ansatz} (\ref{MetricAnsatz}) the Einstein equations (\ref{EinsteinEquations2}) split into the system of the equations for the metric functions:
\begin{eqnarray}\label{SystemOfEquations}
- \frac 32 \dot{u}^2 + \frac 12 S'' - \frac 12 \frac{a}{(1 - a|r|)}S' &=& \frac {1}{M^3}\dot{\varphi}^2~,\nonumber\\
- \frac 32 u'\dot{u} &=& \frac {1}{M^3}\varphi'\dot{\varphi }~,\nonumber\\
(1 - a|r|)\left( u'' - \ddot{u}\right) - a u' &=& 0~,\\
- \frac 32 u'^2 - \frac 12 S'' - \frac 12\frac{a}{( 1 - a|r|)}S' &=& \frac {1}{M^3} \varphi'^2~,\nonumber
\end{eqnarray}
and for the brane tensions:
\begin{eqnarray}\label{Tensions}
M^3\delta (r)\left( \frac 23 a + S'\right) &=& \sigma _{tt} - \frac 13 \frac{e^S}{(1 - a|r|)^{2/3}}\sigma ~, \nonumber \\
M^3\delta (r)\left( \frac 23 a - u'\right)e^{ - S + u} &=& \sigma _{xx} + \frac 13 (1 - a|r|)^{2/3}e^u\sigma ~, \nonumber\\
M^3\delta (r)\left( \frac 23 a - u'\right) e^{ - S + u} &=& \sigma _{yy} + \frac 13 (1 - a|r|)^{2/3}e^u\sigma  ~, \\
M^3\delta (r)\left( \frac 23 a + 2u'\right)e^{ - S - 2u} &=& \sigma _{zz} + \frac 13 (1 - a|r|)^{2/3}e^{-2u}\sigma ~, \nonumber\\
M^3\delta (r)\left( \frac 43 a - S'\right) &=& \sigma _{rr} + \frac 13 \frac{e^S}{(1 - a|r|)^{2/3}}\sigma ~.\nonumber
\end{eqnarray}

The solution to (\ref{BulkScalarFieldEquation}) and (\ref{SystemOfEquations}) is
\begin{eqnarray} \label{SysSolution}
u(t,|r|)&=& A \sin (\omega t) J_0 \left(\frac{\omega }{a} - \omega|r|\right)~,\nonumber\\
\varphi(t,|r|) &=&\sqrt {\frac{3 M^3}{2}} ~A \cos (\omega t) J_0\left(\frac{\omega }{a} - \omega|r|\right)~,\\
S(|r|) &=& \frac {3\omega^2(1-a|r|)^2}{2a^2}A^2\left[ J_0^2\left(\frac{\omega }{a} - \omega|r|\right) + J_1^2\left(\frac{\omega }{a} - \omega|r|\right) - \right. \nonumber \\
&-&\left.\frac {a}{\omega (1-a|r|)} J_0 \left(\frac{\omega }{a} - \omega|r|\right) J_1 \left(\frac{\omega }{a} - \omega|r|\right) \right] - B,\nonumber
\end{eqnarray}
where $A$ and $B$ are some dimensionless constants, $J_0$ and $J_1$ are Bessel functions of the first kind and the integration constant $\omega$ corresponds to the frequency of standing waves.

Using (\ref{SysSolution}) from the equations (\ref{Tensions}) one can easily find the brane energy-momentum tensor:
\begin{equation}\label{SysSolution4}
\sigma _A^B = M^3\delta (r)\mathrm{diag}\left[ \tau _t^t,\tau _x^x,\tau _y^y,\tau _z^z,0 \right]~,
\end{equation}
where the brane tensions are:
\begin{eqnarray}\label{SysSolution5}
\tau _t^t &=& 2a~,\nonumber\\
\tau _x^x &=& \tau _y^y = \frac 23 a + aB + A \omega \sin(\omega t) J_1\left(\frac{\omega }{a}\right) ~,\\
\tau _z^z &=&  \frac 23 a + aB - 2 A \omega \sin(\omega t) J_1\left(\frac{\omega }{a}\right)~.\nonumber
\end{eqnarray}

To interpret the solution (\ref{SysSolution}) as describing the brane at $r=0$, which bounds the scalar and gravitational bulk standing waves, one needs to assume that the oscillatory metric functions and 5D scalar field vanish at the origin, i.e.
\begin{equation}\label{r=0}
u|_{r = 0} = 0~, ~~~ \varphi|_{r = 0} = 0~, ~~~S|_{r = 0} = 0~.
\end{equation}
These conditions can be achieved assuming the relation between $\omega$ and $a$:
\begin{eqnarray}
\label{ConditionsOnTheBrane1}
\frac{\omega }{a} = Z_n^{(J_0)}~,
\end{eqnarray}
where $Z_n^{(J_0)}$ is the $n$-th zero of the function $J_0$, and simultaneously fixing the constant $B$:
\begin{eqnarray}
\label{ConditionsOnTheBrane2}
B = \frac{3\omega^2A^2}{2a^2} J_1^2\left(\frac{\omega}{a}\right)~.
\end{eqnarray}
In fact, the conditions (\ref{r=0}) lead to the quantization, (\ref{ConditionsOnTheBrane1}), of the ratio of the standing wave frequency, $\omega$, to the curvature scale, $a$.

From (\ref{SysSolution}) one can also see that the metric function, $u(t,|r|)$, and the scalar field, $\varphi(t,|r|)$, having similar dependence on spatial coordinates, are oscillating $\pi/2$ out of phase in time, i.e. the energy of the oscillations is passing back and forth between the gravitational and scalar field standing waves bounded by the brane.

There are three free parameters in our model: the constant $A$ (defining the amplitude of oscillations), the curvature scale $a$ (giving the size of extra space for observers on the brane) and the ratio $Z_n^{(J_0)} = \omega/a$ (the $n$-th zero of Bessel function $J_0$). Below we consider two limiting cases corresponding to the small, $A \ll 1$, and the large, $A \gg 1$,  amplitudes of bulk standing waves.


\section{The Small Extra Space}

In the first limiting case the amplitude, $A \ll 1$, and consequently the energy of the oscillations is small. Then, neglecting the terms containing the constants $A$ and $B \sim A^2$ in the brane tensions (\ref{SysSolution5}), one gets the brane energy-momentum tensor,
\begin{equation} \label{sigma}
\sigma _A^B \approx M^3 \delta (r)\mathrm{diag}\left[ 2a,\frac 23 a,\frac 23 a,\frac 23 a,0 \right]~,
\end{equation}
which obeys the equation of state $\cal{E} = \mathrm{3}\cal{P}$, with $\cal{E}$ and $\cal{P}$ being the brane energy density and pressure, respectively.

Moreover, from (\ref{SysSolution}) it is clear that in this case the functions $u$, $S$ and $\varphi$ does not play significant role and from the very beginning one can assume:
\begin{equation}
S(|r|) = u(t,|r|) = \varphi (t,|r|) = 0~,
\end{equation}
and consider the metric {\it ansatz} without oscillatory metric functions:
\begin{equation} \label{MetricAnsatz2}
ds^2 = \frac {1}{(1 - a|r|)^{2/3}}\left( dt^2 - dr^2 \right) - (1 - a|r|)^{2/3}\left( dx^2 + dy^2 + dz^2 \right)~.
\end{equation}

This metric is 5D generalizations of the 4D domain wall solution of \cite{domain-1,domain-2,domain-3}. Due to the presence of the absolute value of the extra coordinate, $|r|$, the Ricci tensor at $r=0$ has $\delta$-function-like singularity which corresponds to the brane tension, see (\ref{Tensions}). The solution (\ref{MetricAnsatz2}) has also new features, since, in contrast to 4D domain walls \cite{domain-1,domain-2,domain-3}, the parameter $a$ has the opposite sign. Due to this fact the metric (\ref{MetricAnsatz2}) has the horizon at $|r| = 1/a$ in the bulk. At these points $R_{tt}$ and $R_{rr}$ components of the Ricci tensor get infinite values, while all gravitational invariants, e.g. Ricci scalar,
\begin{equation}\label{RicciScalar}
R = e^{-S}(1 - a|r|)^{2/3} \left[\frac 32 \left( u'^2 - \dot u^2 \right) + S''\right] - a\left( 2B + \frac 83 \right)\delta (r)~,
\end{equation}
are finite there. It resembles the situation with the Schwarzschild Black Hole, however, in contrast the determinant of our metric (\ref{MetricAnsatz2}) becomes zero at $|r| = 1/a$. As the result nothing can cross the horizon of (\ref{MetricAnsatz2}), and matter fields are confined inside of the 3-brane of the width $\sim 1/a$ in the extra space. To provide experimentally acceptable localization of matter fields the actual size of the extra dimension must be sufficiently small, $\leq 1/M_H$, where $M_H$ denotes the Higgs scale. So in the limiting case $A \ll 1$ the curvature scale $a$ must be large $M_H \leq a \leq M$, where $M$ is the 5D fundamental scale. As regards the brane, its width, in fact, is defined by the horizon size.


\section{The Large Extra Space}

In the second large amplitude limiting case it's obvious that
\begin{eqnarray}\label{brane2}
A^2 \sim B \gg 1~.
\end{eqnarray}
Now, assuming that curvature scale $a$ is relatively small, the width of the brane, located at the origin of the large (of the size $\sim 1/a$) but finite extra space, is determined by the metric function $S (|r|)$ in (\ref{MetricAnsatz}). Indeed, for small $a \sim \omega$ the time-dependent terms in the brane tensions (\ref{SysSolution5}) are negligible,
\begin{equation} \label{sig}
\sigma _A^B \approx M^3\delta (r)\mathrm{diag}\left[ 2a,aB,aB,aB,0 \right]~,
\end{equation}
and the brane width is of the order of $\sim 1/(aB)$. So trapping of matter fields is caused by the pressure of the bulk oscillations and not by the existence of the horizon in the extra space.

As an illustrative example of this trapping mechanism let us consider localization of a real massless scalar field, $\Phi (x^A)$, on the brane in the background metric (\ref{MetricAnsatz}). From the 5D action,
\begin{equation} \label{Sphi}
S_\Phi = \frac 12\int dx^5\sqrt g g^{MN}\partial _M \Phi \partial _N\Phi~,
\end{equation}
we obtain the Klein-Gordon equation for $\Phi (x^A)$:
\begin{equation}\label{ScalFieldEqn}
\frac{1}{\sqrt g} ~\partial_M \left( \sqrt g g^{MN}\partial_N \Phi \right) = 0~.
\end{equation}
Using (\ref{DeterminantOfMetric}) the equation (\ref{ScalFieldEqn}) can be written as:
\begin{eqnarray}\label{ScalFieldEqn1}
\left[ \partial _t^2 - \frac {e^S}{(1 - a|r|)^{4/3}}\left( e^{- u}\partial _x^2 + e^{- u}\partial _y^2 + e^{2u}\partial _z^2 \right) \right]\Phi = \frac {1}{(1 - a|r|)}\partial _r\left[(1 - a|r|)\partial _r \Phi\right]~.
\end{eqnarray}
In addition, it's easy to find that the assumption (\ref{brane2}) leads to the following relation between the metric functions:
\begin{equation}\label{Ratio=S(r)/u(t,r)}
\left|\frac{u(t,|r|)}{S(|r|)}\right| \ll 1~.
\end{equation}
Then in the scalar field equation (\ref{ScalFieldEqn1}) we can drop the function $u(t,r)$ in the exponents and rewrite it as:
\begin{eqnarray}\label{ScalFieldEqn2}
\left[ \partial _t^2 - \frac {e^S}{(1 - a|r|)^{4/3}}\left( \partial _x^2 + \partial _y^2 + \partial _z^2 \right) \right]\Phi = \frac {1}{(1 - a|r|)}\partial _r\left[(1 - a|r|)\partial _r \Phi\right]~.
\end{eqnarray}
We look for the solution in the form:
\begin{equation}\label{Solution1}
\Phi (t,x,y,z,r) = \phi \left( x^\mu \right) \rho (r)~,
\end{equation}
where Greek indexes numerate 4D coordinates, and the 4D factor of the scalar field wavefunction $ \phi \left( x^\mu \right)$ obeys the equation:
\begin{equation}
\eta^{\nu \beta }\partial _\nu \partial _\beta \phi \left( x^\mu \right) = - m^2\phi \left(x^\mu \right)~.
\end{equation}
In what follows we assume
\begin{equation}
\phi \left( x^\mu\right) = e^{ - i\left( Et - p_xx - p_yy - p_zz \right)}~.
\end{equation}
Then extra dimension factor $\rho(\left|r\right|)$ of the scalar field wavefunction obeys the equation:
\begin{equation}\label{GeneralEquation-Rho(r)}
\rho'' - \frac{a ~\mathrm{sgn}( r )}{1 - a|r|}\rho ' - E^2\left[ \frac{e^S}{(1 - a|r|)^{4/3}} - 1 \right]\rho = -m^2\frac{e^S}{(1 - a|r|)^{4/3}}\rho~,
\end{equation}
with the boundary conditions:
\begin{eqnarray}\label{ConditionForRho(r)}
\rho'|_{|r| = 0} = 0~, \nonumber \\
\rho |_{|r| \to 1/a} = 0~.
\end{eqnarray}
For the scalar field zero mode wavefunction, $\rho_0(r)$, with the dispersion relation:
\begin{equation}\label{4D-DispersionRelation}
E^2 = p_x^2 + p_y^2 + p_z^2~,
\end{equation}
the equation (\ref{GeneralEquation-Rho(r)}) reduces to
\begin{equation}\label{Equation-Rho(r)}
\rho_0'' - \frac{a~\mathrm{sgn}}{(1 - a|r|)}\rho_0' - E^2 \left[ \frac{e^S}{(1-a|r|)^{4/3}} - 1 \right] \rho_0 = 0~.
\end{equation}

To show that the equation (\ref{Equation-Rho(r)}) has the solution localized on the brane we investigate the equation in two limiting regions: close to the brane ($|r|\to 0$) and close to the horizon in the extra space ( $|r|\to 1/a$).

In the first limiting region, $|r| \to 0$,
\begin{eqnarray}\label{AtTheOrigin}
u|_{|r| \to 0} &=& \sin (\omega t)\left[ \sqrt{\frac23 B}~a|r| + O\left(a^2|r|^2\right)\right]~, \nonumber \\
S|_{|r| \to 0} &=& - Ba|r| + O\left(a^2|r|^2\right)~,\\
\sqrt g |_{|r| \to 0} &=& 1 - \left( B + \frac 13 \right) a|r| + O \left(a^2|r|^2\right)~,\nonumber
\end{eqnarray}
the equation (\ref{Equation-Rho(r)}) has the following asymptotic form:
\begin{equation}\label{Equation-Rho(r)-AtTheOrigin}
\rho_0'' - a~\mathrm{sgn}(r)\rho_0' + BE^2a|r|\rho_0 = 0~.
\end{equation}
This equation has the general solution:
\begin{eqnarray}\label{SolutionToEquation-Rho(r)-AtTheOrigin}
\rho_0 (r) = e^{a|r|/2}\left[ C_1 Ai\left( \frac 14 \sqrt[3]{\frac{a^4}{B^2E^4}} - \sqrt[3]{\frac{BE^2}{a^2}}a|r| \right) + C_2 Bi\left(\frac14\sqrt[3]{\frac{a^4}{B^2E^4}} - \sqrt[3]{\frac{BE^2}{a^2}}a|r| \right) \right],
\end{eqnarray}
where $C_1$ and $C_2$ are integration constants, and $Ai$ and $Bi$ are Airy functions. To fulfill the conditions (\ref{ConditionForRho(r)}) the constants $C_1$ and $C_2$ must obey the relation:
\begin{equation}\label{RelationBetween-C_1-And-C_2}
C_2 = - \frac{2\sqrt[3]{\frac{BE^2}{a^2}}Ai'\left( \frac14\sqrt[3]{\frac{a^4}{B^2E^4}} \right) - Ai\left( \frac14\sqrt[3]{\frac{a^4}{B^2E^4}} \right)}{2\sqrt[3]{\frac{BE^2}
{a^2}}Bi'\left(\frac14\sqrt[3]{\frac{a^4}{B^2E^4}} \right) - Bi\left( \frac14\sqrt[3]{\frac{a^4}{B^2E^4}} \right)}C_1~,
\end{equation}
where $Ai'$ and $Bi'$ denote the first derivatives of Airy functions. Then at the origin of the extra space, i.e. on the brane, the function $\rho_0 (|r|)$ will have the following series expansion:
\begin{equation}\label{SeriesExpansionOfRhoAtTheOrigin}
\rho_0 (r)|_{|r| \to 0} = C \left( 1 - \frac16 BE^2a|r|^3 \right) + O\left( a^4|r|^4 \right)~,
\end{equation}
where we introduced the constant:
\begin{equation}\label{}
C = 2C_1\sqrt[3]{\frac{BE^2}{a^2}}~\frac{Ai\left( \frac14\sqrt[3]{\frac{a^4}{B^2E^4}}\right) Bi'\left( \frac14 \sqrt[3]{\frac{a^4}{B^2E^4}} \right) - Ai'\left( \frac14\sqrt[3]{\frac{a^4}{B^2E^4}} \right)Bi\left( \frac14\sqrt[3]{\frac{a^4}{B^2E^4}} \right)}{2\sqrt[3]{\frac{BE^2}{a^2}} Bi'\left( \frac14\sqrt[3]{\frac{a^4}{B^2E^4}} \right) - Bi\left( \frac14\sqrt[3]{\frac{a^4}{B^2E^4}} \right)}~.
\end{equation}

In the second limiting region, $|r| \to 1/a$,
\begin{eqnarray}\label{AtTheExtraSpaceEdge}
u|_{|r| \to 1/a} &=& \sin (\omega t)\left[ A - \frac14 A\frac{\omega ^2}{a^2}( 1 - a|r|)^2 + O\left( (1 - a|r|)^3\right) \right]~,\nonumber\\
S|_{|r| \to 1/a} &=&  - B + \frac34 A^2\frac{\omega ^2}{a^2}(1 - a|r|)^2 + O\left( ( 1 - a|r|)^4 \right)~,\\
\sqrt g |_{|r| \to 1/a} &=& e^{-B}(1 - a|r|)^{1/3} + O\left( (1 - a|r|)^{7/3} \right)~,\nonumber
\end{eqnarray}
the equation (\ref{Equation-Rho(r)}) has the following asymptotic form:
\begin{equation}\label{Equation-Rho(r)-AtTheEdges}
\rho_0'' - \frac{a ~\mathrm{sgn}(r)}{1 - a|r|}~\rho_0' - \frac{c^2}{( 1 - a|r|)^{4/3}}~\rho_0 = 0,
\end{equation}
where
\begin{equation}
c^2 =E^2e^{-B}~.
\end{equation}
The general solution to this equation is:
\begin{equation}\label{SolutionToEquation-Rho(xi)-AtTheEdges}
\rho_0 (r) = C_3 I_0\left( \frac {3c}{a}\sqrt[3]{1 - a|r|} \right) + C_4 K_0\left( \frac {3c}{a}\sqrt[3]{1 - a|r|} \right)~,
\end{equation}
where $C_3$ and $C_4$ are some integration constants and $I_0$ and $K_0$ are zero-order modified Bessel functions of the first and second kind, respectively. To fulfill the conditions (\ref{ConditionForRho(r)}) we must choose $C_4=0$ (the function $K_0$ is unbounded function at $|r| \to 1/a$). Then for the series expansion of $\rho_0 (r)$ at $|r| \to 1/a$ we get:
\begin{equation}\label{SeriesExpansionOfRhoAtTheEdge}
\rho_0 (r)|_{|r| \to 1/a} = C_3\left[ 1 + \frac{9c^2}{4a^2}(1 - a|r|)^{2/3} \right] + O\left( (1 - a|r|)^{4/3} \right)~.
\end{equation}

According to (\ref{Ratio=S(r)/u(t,r)}) the action (\ref{Sphi}) of the 5D scalar filed zero mode reduces to:
\begin{equation}\label{L}
S_0 = \frac 12\int dx^4dr\left[ Q_1(r)\partial_t\phi^2 - Q_2(r)\left(\partial_x\phi^2 + \partial_y\phi^2 + \partial_z\phi^2 \right) - Q_3(r) \phi^2 \right]~,
\end{equation}
where integration over $r$ is performed on the finite integration region $\left[-1/a,+1/a\right]$, and $Q_i$ functions are defined as:
\begin{eqnarray}
Q_1(r) &=& (1 - a|r|)\rho_0^2~, \nonumber\\
Q_2(r) &=& (1 - a|r|)^{- 1/3}e^S\rho _0^2~,\\
Q_3(r) &=& (1 - a|r|)\rho_0'^{~2}~.\nonumber
\end{eqnarray}
It is easy to see that for the extra dimension factor of the scalar field zero mode, $\rho_0 (r)$, having asymptotes (\ref{SeriesExpansionOfRhoAtTheOrigin}) and (\ref{SeriesExpansionOfRhoAtTheEdge}), the integral over extra coordinate $r$ in (\ref{L}) is finite. This means that the scalar field zero mode wavefunction is localized on the brane. Also note that on the brane, $r=0$, due to the boundary conditions (\ref{ConditionForRho(r)}), the Lagrangian in the action (\ref{L})  gets the  standard 4D form for the massless scalar field.

To estimate the masses of KK excitations on the brane we consider the scalar particles with zero 3-momentum, $p^2=0$. Then the equation (\ref{GeneralEquation-Rho(r)}) for the massive modes reduces to:
\begin{equation}\label{EquationForMassive-Rho(r)}
\rho'' - \frac{a~ \mathrm{sgn}(r)}{1 - a|r|}\rho' + m^2\rho = 0~.
\end{equation}
The exact general solution to this equation is:
\begin{equation}
\rho (r) = D_1 J_0\left( \frac ma (1 - a|r|) \right) + D_2Y_0\left(\frac ma (1 - a|r|) \right)~,
\end{equation}
where $D_1$ and $D_2$ are some constants. Imposing the boundary conditions (\ref{ConditionForRho(r)}) we get
\begin{eqnarray}
D_2 &=& 0~,\\
J_1\left( \frac ma \right) &=& 0~,
\end{eqnarray}
from which we get the KK mass spectrum of scalar field on the brane:
\begin{equation}
m_n = a~Z_n^{(J_1)}~,
\end{equation}
where $Z_n^{(J_1)}$ is the $n$-th zero of the Bessel function $J_1$. So the mass gap between the zero and the first massive modes will be:
\begin{equation}
\Delta _m = m_1 = a Z_1^{(J_1)} \approx 3.8 a~.
\end{equation}

In the limiting case of this section the curvature scale $a$, which determines the size of extra dimension, is smaller than the scale associated with the width of the brane, $\sim aB$, where $B \gg 1$. So KK modes, having the masses $\approx 3.8a$, can be created in accelerators at relatively low energies, what gives a chance to check this model.


\section{Conclusion}

In this paper we have introduced the new non-stationary 5D braneworld model with the real scalar field in the bulk. The model represents single brane which bounds collective bulk oscillations of gravitational and scalar field standing waves. These waves are out of phase, i.e. the energy of oscillations passes back and forth between the scalar and gravitational waves. The metric of the model has the horizon in the extra space, and consequently, the extra space is finite for the observer on the brane. We have investigated limiting cases of large and small extra space, depending on the two dimensional parameters of the model - the curvature scale and the energy scale associated with the brane.

In the limiting case of small extra space, when the amplitude of the standing waves is small, these two parameters are of a same order of magnitude, and matter fields are localized on the brane due to the presence of the metric horizon.

In the case of large oscillations the distance to the horizon is relatively large as compared with the brane width, and trapping of matter fields on the brane is caused mainly by the pressure of bulk oscillations. The mass of the lightest KK mode in this case is determined by the smaller energy scale associated with the horizon size, and therefore such particles can be created in future accelerators at relatively low energies.


\section*{Acknowledgments}

MG was partially supported by the grant of Shota Rustaveli National Science Foundation $\#{\rm DI}/8/6-100/12$. The research of PM was supported by Ilia State University.


\end{document}